\documentclass[11pt]{article}
\usepackage{epsfig,graphicx,cite}

\input amssym.def
\input amssym.tex

\makeatletter
\@addtoreset{equation}{section}
\makeatother

\def\ben{\begin{equation}}
\def\een{\end{equation}}

\def\bea{\begin{eqnarray}}
\def\eea{\end{eqnarray}}

\def\CP{{{\mathbb C}{\mathbb P}}}
\def\wR{{\widehat R}}

\def\aa{\mu}
\def\bb{\nu}
\def\cc{\rho}
\def\dd{\sigma}

\def\mathbb{\Bbb}
\def\be{\begin{equation}}
\def\ee{\end{equation}}
\def\nn{\nonumber}
\newcount\hour \newcount\minute
\hour=\time  \divide \hour by 60
\minute=\time

\def\mathbb{\Bbb}

\def\ft#1#2{{\textstyle{\frac{\scriptstyle #1}{\scriptstyle #2} } }}
\def\fft#1#2{{\frac{#1}{#2}}}

\newcount\hour \newcount\minute
\hour=\time  \divide \hour by 60
\minute=\time
\loop \ifnum \minute > 59 \advance \minute by -60 \repeat
\def\nowtwelve{\ifnum \hour<13 \number\hour:
                      \ifnum \minute<10 0\fi
                      \number\minute
                      \ifnum \hour<12 \ A.M.\else \ P.M.\fi
         \else \advance \hour by -12 \number\hour:
                      \ifnum \minute<10 0\fi
                      \number\minute \ P.M.\fi}
\def\nowtwentyfour{\ifnum \hour<10 0\fi
                \number\hour:
                \ifnum \minute<10 0\fi
                \number\minute}
\def\now{\nowtwelve}
\newcommand{\hoch}[1]{$\, ^{#1}$}

\newcommand{\auth}{\Large\bf{M. Cveti\v c\hoch{1,2,5},
Xing-Hui Feng\hoch{1}, H. L\"u\hoch{1} and C.N. Pope\hoch{1,3,4}
}}

\begin{document}
\begin{flushright}
\hfill {UPR-1282-T\ \ \ MI-TH-1629
}\\
\end{flushright}

\begin{center}

{\LARGE{\bf Rotating Solutions in Critical Lovelock Gravities}}

\vspace{9pt}
\auth

\vspace{10pt}

{\small

{\hoch{1}\it Department of Physics, Beijing Normal University, Beijing,
100875,
China}

\vspace{0pt}{\hoch{2}\it Department of Physics and Astronomy,\\
University of Pennsylvania, Philadelphia, PA 19104, USA}

\vspace{0pt}{\hoch{3}\it Mitchell
Institute for Fundamental
Physics and Astronomy,\\
Texas A\&M University, College Station, TX 77843-4242, USA
}

\vspace{0pt}{\hoch{4}\it DAMTP, Centre for Mathematical Sciences,\\
 Cambridge University, Wilberforce Road, Cambridge CB3 OWA, UK}

\vspace{0pt}{\hoch{5}\it Center for Applied Mathematics and Theoretical Physics,\\
University of Maribor, SI2000 Maribor, Slovenia}

}

\newcount\hour \newcount\minute
\hour=\time  \divide \hour by 60
\minute=\time
\loop \ifnum \minute > 59 \advance \minute by -60 \repeat
\def\nowtwelve{\ifnum \hour<13 \number\hour:
                      \ifnum \minute<10 0\fi
                      \number\minute
                      \ifnum \hour<12 \ A.M.\else \ P.M.\fi
         \else \advance \hour by -12 \number\hour:
                      \ifnum \minute<10 0\fi
                      \number\minute \ P.M.\fi}
\def\nowtwentyfour{\ifnum \hour<10 0\fi
                \number\hour:
                \ifnum \minute<10 0\fi
                \number\minute}
\def\now{\nowtwelve}


%
\vspace{40pt}

\underline{ABSTRACT}
\end{center}

 For appropriate choices of the coupling constants, the equations of motion
of Lovelock gravities up to order $n$ in the Riemann tensor can be
factorized such that the theories admits a single (A)dS vacuum.  In this
paper we construct two classes of exact rotating metrics in such critical
Lovelock gravities of order $n$ in $d=2n+1$ dimensions.  In one class,
the $n$ angular momenta in the $n$ orthogonal spatial 2-planes are equal,
and hence the metric is of cohomogeneity one.  We construct these metrics
in a Kerr-Schild form,  but they can then be recast in terms of
Boyer-Lindquist coordinates.  The other class involves metrics with
only a single non-vanishing angular momentum.  Again we construct them
in a Kerr-Schild form, but in this case it does not seem to be
possible to recast them in Boyer-Lindquist form.  Both classes of
solutions have naked curvature singularities, arising because of the
over rotation of the configurations.

%

\vfill
{\footnotesize cvetic@hep.upenn.edu\ \ \ xhfengp@mail.bnu.edu.cn\ \ \
mrhonglu@gmail.com\ \ \ pope@physics.tamu.edu}

\pagebreak

\section{Introduction}

    Although Einstein's theory of gravity is highly non-linear, exact
solutions do exist, including the celebrated (static)
Schwarzschild \cite{Schwarzschild:1916uq} and (rotating)
Kerr \cite{Kerr:1963ud} metrics.  Whilst the generalization of the
Schwarzschild metric to higher dimensions is straightforward, such a
generalization of the Kerr metric leads to still exact, but considerably
more complicated solutions \cite{Myers:1986un}, especially so when the
metrics are asymptotic to (A)dS ((anti-)de Sitter) spacetimes \cite{Hawking:1998kw,Gibbons:2004uw,Gibbons:2004js}.

Finding exact solutions becomes much more difficult when Einstein gravity
is extended with higher-order curvature invariants, even in the
case of static solutions.  Einstein gravity extended with quadratic
curvature invariants in four dimensions was shown by numerical methods to
admit a new static black hole over and above the Schwarzschild metric,
but no exact solution is known \cite{Lu:2015cqa,Lu:2015psa}.  The existence
of such new black holes was shown numerically also when a cosmological
constant or a Maxwell field is included \cite{Lin:2016jjl,Lin:2016kip}.

In higher dimensions, when higher-order ghost-free Euler integrands are no
longer total derivatives, Einstein-Gauss-Bonnet or more general Lovelock
gravities can be constructed \cite{Lovelock:1971yv}.  In these theories,
exact solutions for static black holes have been found
\cite{Boulware:1985wk,Cai:2001dz}, and these have smooth limits to the
Schwarzschild metric when the higher-derivative couplings are sent to zero.
Exact solutions for rotating black holes remain elusive in these theories.

Recently, a five-dimensional rotating solution \cite{Anabalon:2009kq} was
constructed in the Einstein-Gauss-Bonnet (EGB) theory, for a certain
critical value of the coupling constant for the Gauss-Bonnet term.
For generic values of the coupling constant, the EGB theory admits
two (A)dS vacua with different cosmological constants.  One of these
has a positive kinetic energy for linearized graviton fluctuations, while
the other has a negative kinetic energy \cite{Boulware:1985wk}. At the
critical value of the coupling, the two values for the (A)dS cosmological
constants coalesce, and the linearized equations of motion are
automatically satisfied, leading to a gravity theory without a
linearized graviton fluctuation \cite{Fan:2016zfs}, and for which
further exact solutions can be constructed.

The equations of motion of for higher-order Lovelock gravities can
also be factorized for  certain specific choices of the coupling
constants, again giving rise to only a single (A)dS vacuum with one
specific cosmological constant.  Such theories were classified and
studied in \cite{Crisostomo:2000bb}. The critical EGB theory mentioned
earlier is a special case.  The purpose of this paper is to generalize the
five dimensional rotating solution that was
found in  \cite{Anabalon:2009kq} for the critical EGB theory to the critical
Lovelock gravities of order $n$ in the Riemann tensor, in the spacetime
dimension $d=2n+1$. We obtain exact rotating solutions in two cases.
In the first, the $n$ angular momenta in the $n$ orthogonal spatial 2-planes
are all equal, and hence the metric is of cohomogeneity one.  We obtain these
solutions first in a Kerr-Schild form, but we find that
they can then be recast into a form written using Boyer-Linquist type
coordinates. This rewriting has the advantage that it is easier to
study the global structure of the solutions.  The second class of
rotating solutions that we obtain involve only a single
non-vanishing angular momentum.  Again, we obtain the solutions in
a Kerr-Schild form, but in this case there appears to be no way
to introduce Boyer-Lindquist type coordinates.

The paper is organized as follows.  In section 2, we review the
construction of the critical Lovelock gravities. In section 3, we
consider static and spherically-symmetric solutions.  Next, we focus on
Lovelock gravities of order $n$ in $d=2n+1$ dimensions.  In section 4, we
construct the exact rotating solutions where all the angular momenta
are equal.  In section 5, we construct the second class of rotating
solutions, where only a single angular momentum is non-zero.  We conclude
the paper in section 6.  In the appendix, we present details of
the Riemann tensor for the single-angular momentum metrics.

\section{Critical Lovelock gravities}

In this section, we review the construction of \cite{Crisostomo:2000bb}.
We start with the general class of Lovelock gravities, for which the
Lagrangian is given by
\be
e^{-1} {\cal L} = \sum_{k=0}^n \alpha_k E^{(k)}\,,\label{lovelocklag}
\ee
where
\be
E^{(k)}=\fft{1}{2^{k}} \delta_{\aa_1\bb_1\cdots \aa_k \bb_k}^{
 \cc_1\dd_1\cdots \cc_k \dd_k}
R^{\aa_1\bb_1}_{\cc_1\dd_1}\cdots R^{\aa_k\bb_k}_{\cc_k\dd_k}\,,
\ee
and $R^{\aa\bb}_{\cc\dd}$ denotes the Riemann tensor
$R^{\aa\bb}{}_{\cc\dd}$
and\footnote{Note that this normalisation for
$\delta_{\alpha_1\cdots\alpha_{s}}^{\beta_1\cdots\beta_{s}}$ is not the
rather standard ``unit-strength'' convention.}
\be
\delta_{\alpha_1\cdots\alpha_{s}}^{\beta_1\cdots\beta_{s}}=
s! \delta_{[\alpha_1}^{\beta_1} \cdots \delta_{\alpha_s]}^{\beta_s}\,.
\ee
The Euler integrands $E^{(k)}$ can also be expressed as
\be
E^{(k)} = \ft{(2k)!}{2^k}\, R^{[\aa_1\bb_1}_{\aa_1\bb_1}
\cdots R^{\aa_k \bb_k]}_{\aa_k \bb_k}\,.
\ee
The first few cases are given by
\be
E^{(0)} =1\,,\qquad
E^{(1)}=R\,,\qquad E^{(2)} = R^2 - 4R^{\mu\nu}R_{\mu\nu} +
R^{\mu\nu\rho\sigma}R_{\mu\nu\rho\sigma}\,,\quad \hbox{etc.}
\ee
In order for all the Euler integrands $E^{(k)}$ in (\ref{lovelocklag})
to be non-trivial, the spacetime dimension $d$ should be $\ge 2n+1$.

The term $\sqrt{-g} E^{(k)}$ in the Lagrangian (\ref{lovelocklag})
gives a contribution
\be
E_{\mu}^{(k)\,\nu} = -\ft{1}{2^{k+1}}
\delta^{\cc_1\dd_1\cdots \cc_k \dd_k\,\nu}_{\aa_1\bb_1\cdots \aa_k \bb_k\,\mu}\,
R^{\aa_1\bb_1}_{\cc_1\dd_1}\cdots R^{\aa_k \bb_k}_{\cc_k \dd_k}
\ee
to the equations of motion.

The equations of motion following from
(\ref{lovelocklag}) imply that  general the condition for
an (A)dS spacetime
with $R_{\mu\nu}= \lambda\, g_{\mu\nu}$ to be a solution is that $\lambda$
should be any of the roots of a
certain $n$'th -rder polynomial, with coefficients
proportional to the constants $\alpha_k$.  By choosing the coefficients
$\alpha_k$ appropriately, one can arrange that all the roots are equal.
This case corresponds to having the equations of motion
\be
E_\mu^\nu\equiv -\ft{1}{2^{n+1}}
\delta^{\cc_1\dd_1\cdots \cc_n \dd_n\,\nu}_{\aa_1\bb_1\cdots \aa_n \bb_n\,\mu}\,
\widehat R^{\aa_1\bb_1}_{\cc_1\dd_1}\cdots
  \widehat R^{\aa_n \bb_n}_{\cc_n \dd_n}=0\,.
\label{geneoms}
\ee
where $\widehat R^{\mu\nu}_{\rho\sigma}$, which we shall refer to as
the subtracted Riemann tensor,\footnote{Note that in the case of
an Einstein metric with cosmological constant such that $R_{\mu\nu}=
 -(n-1)\, \ell^{-2}\, g_{\mu\nu}$, the subtracted Riemann tensor
(\ref{hatRdef}) is nothing but the Weyl tensor.} is
given by
\be
\widehat R^{\aa\bb}_{\cc\dd} = R^{\aa\bb}_{\cc\dd} +
  \fft{1}{\ell^2} \delta^{\aa\bb}_{\cc\dd}\,.
\label{hatRdef}
\ee
The subtracted Riemann tensor vanishes in the case of an AdS vacuum
with radius $\ell$.  We could, alternatively, obtain a de Sitter solution,
by taking $\ell^2<0$. It turns out that
\bea
E_{\mu}^\nu &=& \sum_{k=0}^n \big(\fft{2}{\ell^2}\big)^{n-k}
C_n^k \fft{(d-2k-1)!}{(d-2n-1)!}\fft{2^{k+1}}{2^{n+1}} E_{\mu}^{(k)\nu}\,,\cr
&=& \sum_{k=0}^n \big(\fft{1}{\ell^2})^{n-k}\, C_n^k\, \fft{(d-2k-1)!}{(d-2n-1)!}\,
E_\mu^{(k)\nu}\,.
\eea
These theories were constructed and studied in \cite{Crisostomo:2000bb}.  We
shall refer to them as critical Lovelock gravities of order $n$.

In this paper, we are interested in the case with $d=2n+1$, corresponding
to the critical gravity of maximum order in a given odd dimension. Thus we
have
\be
E_{\mu}^\nu = \sum_{k=0}^n \big(\fft{1}{\ell^2})^{n-k}\, C_n^k\, (2(n-k))!\,
E_\mu^{(k)\nu}\,.
\ee
The corresponding Lagrangian is thus
\be
e^{-1} {\cal L} = \sum_{k=0}^n \big(\fft{1}{\ell^2})^{n-k}\, C_n^k\, (2(n-k))!\,
E^{(k)}\,.
\ee

The critical Lovelock gravities are characterized by the fact that they
admit only a single (A)dS vacuum, for which the subtracted Riemann tensor
vanishes.  The linearization around the AdS vacuum was studied
for the five-dimensional case (i.e.~Einstein-Gauss-Bonnet, EGB)
in \cite{Fan:2016zfs}.  It turns out
that for a generic EGB theory, where there are two inequivalent
(A)dS vacua, the kinetic term for the linearized graviton gives positive
energy in one vacuum, and negative energy in the other.  When the two (A)dS
spacetimes coalesce, i.e.~in the critical theory, the linearized perturbation
equations become vacuous.  The perturbation equations at quadratic
order were derived in \cite{Fan:2016zfs}. It is straightforward to see that
for the critical Lovelock gravity of order $n$, the analogous
perturbation equations up to and including order $(n-1)$ are vacuous.

\section{Static solutions}

In this paper, we are interested in constructing solutions where, unlike
in (A)dS, the subtracted Riemann tensor does not vanish. The simplest
such case is perhaps a static, spherically-symmetric metric, for which
the most
general ansatz takes the form
\be
ds^2=-h(r) dt^2 + \fft{dr^2}{f(r)} + r^2 d\Omega_{d-2,\epsilon}^2\,,\qquad
d\Omega_{d-2,\epsilon}^2 = \fft{dy^2}{1-\epsilon y^2} + y^2 d\Omega_{d-3}^2\,,
\ee
with $\epsilon=1,0,-1$, and $d\Omega_{d-3}^2$ is the metric for a unit
round $S^{d-3}$.  (To be precise, we include the topologies $T^{d-2}$
and $H^{d-2}$ also, corresponding to taking $\epsilon=0,-1$ respectively.)
The critical theories admit black hole solutions with $h=f$.  These solutions
were obtained in \cite{Crisostomo:2000bb}. For $d=2n+1$, the solution
becomes particular simple, being given by
\be
h=f=r^2/\ell^2 + \epsilon - \mu\,,\label{static}
\ee
where $\mu$ is an integration constant.

It was shown in \cite{Fan:2016zfs} that the critical EGB theory admits
another type of static solution, with
\be
f=r^2/\ell^2 + \epsilon\,,\qquad
h=h(r)\,\,\, \hbox{is an arbitrary function.}\label{statsol}
\ee
We may easily check that in fact this gives a solution in
all the critical Lovelock gravities:
The subtracted Riemann tensor $\wR^{\mu\nu}_{\rho\sigma}$ is given by
\be
\widehat R^{ti}_{tj} = \Big(\fft{1}{\ell^2} - \fft{(\epsilon + r^2/\ell^2)h'}{2rh}\Big)\delta^{i}_j\,,\qquad
\widehat R^{tr}_{tr}=\fft{rh'-2h}{2h\ell^2} -
\fft{(2hh''-h'^2)(\epsilon+r^2/\ell^2)}{4h^2}\,,\label{staticR}
\ee
with all remaining components, aside from those following from (\ref{staticR})
by the Riemann tensor symmetries, vanishing identically.  As expected,
the subtracted Riemann tensor vanishes when $h=r^2/\ell^2 + \epsilon$,
corresponding to the AdS vacuum.
It is straightforward to see that (\ref{staticR}) satisfies the
equations of motion (\ref{geneoms}) in all the critical
Lovelock theories in $d\ge 2n+1$ dimensions, since the non-vanishing
components of the subtracted Riemann tensor are not sufficient to span the
entire range of index values required by the
antisymmetric $\delta$-tensor in (\ref{geneoms}).

\section{Rotating solutions: all equal rotation}

   A rotating solution in the five-dimensional critical EGB theory
was obtained in \cite{Anabalon:2009kq}, by taking the metric to have
a Kerr-Schild form with an AdS ``base'' $d\bar s^2$ that is written in
spheroidal
coordinates.  The geodesic null vector $K_\mu$
that is used in the Kerr-Schild
construction in \cite{Anabalon:2009kq} is the same as the one used in
the construction of the
Kerr-Schild form of the five-dimensional Kerr-AdS metric in
\cite{Gibbons:2004uw,Gibbons:2004js}.  However, the function $w$ in the
Kerr-Schild metric $ds^2= d\bar s^2 + w\, (K_\mu dx^\mu)^2$ is quite different
in \cite{Anabalon:2009kq} from the one in \cite{Gibbons:2004uw,Gibbons:2004js}
that gives Kerr-AdS.

   The rotating solution in \cite{Anabalon:2009kq} has independent rotation
parameters in the two orthogonal spatial 2-planes.  We have looked
without success for
analogous solutions with independent rotation parameters in the
higher-dimensional critical Lovelock gravities (\ref{geneoms})
in dimensions $d=2n+1$.
However, we have been able to construct higher-dimensional generalisations
in the case where all the rotation parameters are taken to be equal.
As was shown in \cite{Gibbons:2004uw,Gibbons:2004js}, the AdS base metric
can then be written in terms of the Fubini-Study metric on $\CP^{n-1}$.
We find that the full Kerr-Schild metric for the critical Lovelock
solution takes the form
\be
ds^2 = d\bar s^2 + \lambda (r^2 + a^2)  K^2\,,\label{KSmet}
\ee
with
\bea
d\bar s^2 &=& - \fft{(g^2r^2 + 1) dt^2}{\Xi} +
\fft{r^2 dr^2}{(g^2 r^2 + 1) (r^2 + a^2)} +
\fft{r^2 + a^2}{\Xi} \big[(d\psi + A)^2 + d\Sigma_{n-1}^2\big]\,,\cr
K &=& K_\mu dx^\mu = \fft{1}{\Xi} \big[dt - a (d\psi + A)\big] +
\fft{r^2 dr}{(g^2 r^2 + 1) (r^2 + a^2)}\,,
\eea
where $\Xi=1-g^2 a^2$ and $d\Sigma_{n-1}^2$ is the Fubini-Study metric on $\CP^{n-1}$, with
its canonical normalisation $\bar R_{ab}= 2n\, \bar g_{ab}$.  As with the
Kerr-AdS metrics in odd dimension and with equal angular momenta, the
solutions we obtain here have cohomogeneity one.

The metric (\ref{KSmet}) can be recast in terms of
Boyer-Lindquist coordinates (for which there are
no cross terms between $dr$ and the other coordinate differentials), by
means of the transformations
\bea
dt &=& d\tau + \fft{\lambda r^2\, dr}{(1+g^2 r^2)(1-(\lambda-g^2) r^2)}
\,,\cr
d\psi &=& d\sigma + g^2 a\, d\tau +
\fft{a \lambda r^2\, dr}{(r^2+a^2)(1-(\lambda-g^2) r^2)} \,.
\eea
The metric (\ref{KSmet}) then becomes
\bea
ds^2= -\fft{\rho^2\, h^2}{r^2}\, \Big(d\tau - \fft{a}{\Xi}\,
(d\sigma + A)\Big)^2  + \fft{\rho^4}{r^2}\, \Big( \fft{(d\sigma + A)}{\Xi}
  -\fft{a}{\rho^2}\, d\tau\Big)^2 + \fft{d\rho^2}{h^2} + \fft{\rho^2}{\Xi}\, d\Sigma_{n-1}^2\,,\label{BLmet}
\eea
where we use $\rho=\sqrt{r^2+a^2}$ as the radial variable, and
\be
r^2=\rho^2=a^2\,,\qquad h^2 = 1 - (\lambda-g^2)(\rho^2-a^2)\,.
\label{hdef}
\ee

We now prove that the metric (\ref{BLmet}) indeed
satisfies the equation (\ref{geneoms}). It is convenient to define the
vielbein basis
\bea
e^0 &=& \fft{\rho h}{r}\,  \Big(d\tau - \fft{a}{\Xi}\,
(d\sigma + A)\Big)\,,\qquad e^1= \fft{d\rho}{h}\,,\cr
e^2&=& \fft{\rho^2}{r}\, \Big( \fft{(d\tau + A)}{\Xi}
  -\fft{a}{\rho^2}\, d\tau\Big)\,,\qquad
  e^a = \fft{\rho}{\sqrt{\Xi}}\, \bar e^a\,,
\eea
where $\bar e^a$ is a vielbein basis for $\CP^{n-1}$.  In fact, for the
purposes of the calculations below, we need not restrict the
metric $d\Sigma_{n-1}^2$ to be that of $\CP^{n-1}$ specifically; we
may take it to be any K\"ahler metric on a complex
manifold ${\cal K}^{n-1}$ of complex dimension $n-1$.

   With the function $h(\rho)$ as yet arbitrary, the torsion-free spin
connection is given by
\bea
\omega_{01}&=& -\fft{r}{\rho}\, \bigg(\fft{\rho h}{r}\bigg)' \, e^0
  + \fft{a}{r^2}\, e^2\,,\quad \omega_{02}= \fft{a}{r^2}\, e^1\,,\quad
  \omega_{0a}= \fft{h a}{\rho r}\, J_{ab}\, e^b\,,\quad
\omega_{1a} = -\fft{h}{\rho}\, e^a\,,\\
\omega_{12} &=& \bigg( \fft{h r'}{r} - \fft{2 h \rho}{r^2}\bigg)\, e^2 -
   \fft{a}{r^2}\, e^0\,,\quad \omega_{2a} = \fft1{r}\, J_{ab}
\, e^b\,,\quad \omega_{ab} = \bar\omega_{ab} -
  \fft{h a}{\rho r}\, J_{ab}\, e^0 - \fft1{r}\, J_{ab}\, e^2\,,\nn
\eea
where a prime denotes a derivative with respect to $\rho$,
$J_{ab}$ are the vielbein components of the Ka\"ahler form
of ${\cal K}^{n-1}$, i.e.\ $J=\ft12 J_{ab}\, \bar e^a\wedge \bar e^b$, and
$\bar \omega_{ab}$ is the spin connection of ${\cal K}^{n-1}$.
The curvature 2-forms, after taking $h(\rho)$ to be given by (\ref{hdef}),
are given by
\bea
\Theta_{01} &=& -(\lambda-g^2)\, e^0\wedge e^1\,,\qquad \Theta_{02} =
   -(\lambda-g^2)\, e^0\wedge e^2\,,\nn\\
\Theta_{0a} &=& -(\lambda-g^2)\, e^0\wedge e^a +
 \fft{h a \sqrt{\Xi}}{\rho^2\, r}\, (\bar\nabla_c\, J_{ab}) \,
 e^c\wedge e^b\,,\qquad
\Theta_{1a}= (\lambda-g^2)\, e^1\wedge e^a\,,\nn\\
\Theta_{2a} &=& (\lambda-g^2)\, e^2\wedge e^a + \fft{\sqrt{\Xi}}{\rho r}\,
   (\bar\nabla_c\, J_{ab}) \,
 e^c\wedge e^b\,,\nn\\
\Theta_{ab} &=& \bar\Theta_{ab} +  (\lambda-g^2)\, e^a\wedge e^b -
 \fft{1+a^2(\lambda-g^2)}{2\rho^2}\, \Omega_{abcd}\, e^c\wedge e^d\nn\\
&& -
 (\bar\nabla_c\, J_{ab}) \, \bigg( \fft{h a \sqrt{\Xi}}{\rho^2\, r} \,
    e^c\wedge e^0 + \fft1{r}\, e^c\wedge e^2\bigg)\,,\label{Thetaab}
\eea
where $\bar\Theta_{ab}$ is the curvature 2-form of ${\cal K}^{n-1}$, and
\be
\Omega_{abcd}= \delta_{ac}\, \delta_{bd} - \delta_{ad}\, \delta_{bc} +
  J_{ac}\, J_{bd}- J_{ad}\, J_{bc} + 2 J_{ab}\, J_{cd}\,.
\ee
Note that all the terms involving
$(\bar\nabla_c\, J_{ab})$ in (\ref{Thetaab}) actually vanish,
since the K\"ahler form is covariantly constant.

  It is now evident that if we define $\widehat R^{\mu\nu}{}_{\rho\sigma}$ as
in (\ref{hatRdef}),
then provided we choose $g$ and
$\lambda$ such that
\be
\fft{1}{\ell^2} = g^2 - \lambda\,,\label{effeom}
\ee
then the only non-vanishing components of $\widehat R^{\mu\nu}{}_{\rho\sigma}$
will be when the indices lie in the directions of the K\"ahler manifold,
with
\be
\widehat R^{ab}_{cd} = \fft{\Xi}{\rho^2}\, \bar R^{ab}{}_{cd}
    -\fft{1 + a^2(\lambda-g^2)}{\rho^2}\, \Omega^{ab}{}_{cd}\,.
\ee
Since these non-zero components lie within a $(2n-2)$-dimensional subspace
of the full $(2n+1)$-dimensional spacetime, it follows that the
antisymmetrisations in (\ref{geneoms}) will ensure that the field equations
are satisfied.

   Note that this gives a solution of the equations of motion when
${\cal K}^{n-1}$ is {\it any} K\"ahler manifold.  For the particular
case we started with, when ${\cal K}^{n-1}= \CP^{n-1}$ with its standard
Fubini-Study metric which has constant holomorphic sectional curvture,
\be
\bar R_{abcd}= \Omega_{abcd}\,,
\ee
we have the especially simple result that
\be
\widehat R^{ab}_{cd} = -\fft{a^2\, \lambda}{\rho^2} \, \Omega^{ab}{}_{cd}\,.
\ee

\section{Rotating solutions: a single rotation}

  We have also been able to construct rotating solutions in the
$d=2n+1$ dimensional critical Lovelock gravities (\ref{geneoms}) in
tha case that just a single rotation parameter is non-vanishing.
The metric in $d=2n+1$ dimensions is given by
\be
ds^2 = d\bar s^2 + \lambda\, \rho^2  K^2\,,\label{1rot1}
\ee
with
\bea
d\bar s^2 &=& - \fft{(g^2r^2 + 1) \Delta_\theta dt^2}{1-a^2 g^2} +
\fft{\rho^2 dr^2}{(g^2r^2 + 1)(r^2 + a^2)} +
\fft{\rho^2 d\theta^2}{\Delta_\theta}\nn\\
&+& \fft{(r^2 + a^2) \sin^2\theta d\phi^2}{1 -a^2 g^2} +
r^2 \cos^2\theta\, d\Omega_{2n-3}^2\,,\nn\\
K &=& K_\mu dx^\mu = \fft{\Delta_\theta\, dt}{1-a^2 g^2} -
\fft{\rho^2 dr}{(g^2 r^2 + 1) (r^2 + a^2)} -
\fft{a \sin^2\theta d\phi}{1 - a^2 g^2}\,,\nn\\
\rho^2 &=& r^2 + a^2 \cos^2\theta\,,\qquad
\Delta_\theta =1 - a^2 g^2 \cos^2\theta\,.\label{1rot2}
\eea
If we choose $g$ and $\lambda$ to satisfy (\ref{effeom}), then we find that
the non-vanishing components of the subtracted Riemann tensor
$\hat R^{\mu\nu}_{\rho\sigma}$, defined in (\ref{hatRdef}), are given
by the expressions in appendix A.
Decomposing the indices as $\mu=(a,i)$, etc.,\
where $x^a=(t,r,\theta,\phi)$ and $x^i$ are the coordinates of the
$(2n-3)$-sphere, the
non-vanishing components of $\hat R^{\mu\nu}_{\rho\sigma}$ are
of the forms
\be
\hat R^{ab}_{cd}\,,\qquad \hat R^{ai}_{bj} = T^a_b\, \delta^i_j\,,\qquad
\hat R^{ij}_{k\ell} = f\,
   \delta^{ij}_{k\ell}\,.\label{hatRforms}
\ee
The expressions for $f$, $T^a_b$ and $\wR^{ab}_{cd}$ can be found in
(\ref{fres}), (\ref{Tab}) and (\ref{Rabcd}) respectively.
A crucial point for what follows is that the expressions for the
components of $\hat R^{\mu\nu}_{\rho\sigma}$ are completely independent
of the spacetime dimension (except for the obvious fact that the
range of the $i$ index is dimension dependent).  Furthermore, the
non-vanishing components of $\wR^{\mu\nu}_{\rho\sigma}$ have either
four, two (one up, one down) or zero $(2n-3)$-sphere indices.

  Given the structure of the non-vanishing components of
$\wR^{\mu\nu}_{\rho\sigma}$, it is clear from (\ref{geneoms}) that
the only non-trivial equations of motion will be
\be
E_a^b=0 \quad \hbox{and} \quad E_i^j=0\,.
\ee
Furthermore, we see that
\bea
E_a^b &=& \alpha_1 \, f^{n-2}\, S^{(1)\, b}_a +
            \alpha_2\, f^{n-3}\, S^{(3)\, b}_a\,,\nn\\
E_i^j &=& \alpha_3\, f^{n-2}\, S^{(0)}\, \delta_i^j +
   \alpha_4\, f^{n-3}\, S^{(1)\, b}_a\, T^a_b\, \delta_i^j +
   \alpha_5\, f^{n-4}\, S^{(3)\, b}_a\, T^a_b\, \delta_i^j\,,
\eea
where the $\alpha$ coefficients are non-vanishing combinatoric factors,
and
\be
S^{(0)}= \delta^{c_1 d_1 c_2 d_2}_{a_1 b_1 a_2 b_2}\,
   \wR^{a_1 b_1}_{c_1 d_1}\, \wR^{a_2 b_2}_{c_2 d_2}\,,\quad
S^{(1)\, b}_a = \delta^{b c_1 d_1 c_2}_{a a_1 b_1 a_2}\,
            \wR^{a_1 b_1}_{c_1 d_1}\, T^{a_2}_{c_2}\,,\quad
S^{(3)\, b}_a =\delta^{b d_1 c_2 d_2}_{a b_1 a_2 b_2}\,
                T^{a_1}_{c_1}\, T^{b_1}_{d_1}\, T^{a_2}_{c_2}\,.
\ee
After rather intricate, but mechanical calculations (which we performed
using Mathematica), we find that
\be
S^{(0)}=0\,,\qquad S^{(1)\, b}_a=0\,,\qquad S^{(3)\, b}_a=0\,,
\ee
and hence the single rotation metrics satisfy the equations of
motion (\ref{geneoms}) in all dimensions $d=2n+1$, provided that
(\ref{effeom}) holds.

\section{Conclusions}

In this paper, we considered critical Lovelock gravities and focused on
those of order $n$ in $d=2n+1$ dimensions.  We obtained two classes of
rotating solutions.  In the first class, all the angular momentum parameters
are set equal, and the metric is of cohomogeneity one.  We presented the
metric in both the Kerr-Schild and Boyer-Lindquist forms. In the
second class of solutions, only a single rotation parameter is non-vanishing,
and the solution is obtained in the Kerr-Schild form.  In this case,
it does not appear to be possible to rewrite it in terms of Boyer-Lindquist type
coordinates.  By calculating the subtracted Riemann tensor that appears
in the equations of motion (\ref{geneoms}) explicitly, we
demonstrated that the metrics in both of the classes indeed satisfy
the equations of motion.  When restricted to five dimensions, our solutions
are special cases of the rotating solutions constructed in
\cite{Anabalon:2009kq}.

The metrics are all asymptotic to AdS, but they do not describe black holes.
Rather, they have naked curvature singularities. The analysis is rather
straightforward for the solution with where all the angular momenta
are equal, since in this case we can rewrite the metric using
Boyer-Lindquist type coordinates.  Another way to see the geometric
structure  by noting that if we set the rotation parameter to zero,
the solution reduces to AdS, with no ``mass'' parameter analogous
to $\mu$ in the static solutions (\ref{static}). The naked singularity
can thus be understood as being associated with a solution that
is ``over rotated,'' in the sense that it has angular momentum but no mass.

Exact rotating solutions are hard to come by, and although the
solutions we have obtained here have shortcomings associated with the
presence of naked singularities, they do perhaps provide a guide as
to how one might hope to construct more general rotating solutions in
critical Lovelock gravities.  It would be of great interest to try to
obtain such generalisations where a mass parameter could be added,
so that rotating black hole solutions without naked singularities
might be possible.  It would also be interesting to seek
rotating solutions in the higher-dimensional critical Lovelock
gravities in which the angular momentum parameters could be arbitrary.

\section*{Acknowlegement}

M.C.~research is supported in part by the DOE Grant Award DE-SC0013528,
the Fay R. and Eugene L. Langberg Endowed Chair  and the Slovenian
Research Agency (ARRS). X.-H.F.~and H.L.~are supported in part by NSFC
grants NO. 11175269, NO. 11475024 and NO. 11235003. C.N.P.~is supported
in part by DOE grant DE-FG02-13ER42020.

\appendix

\section{Subtracted Riemann Tensor for Single-Rotation Metrics}

The components of the subtracted Riemann tensor (\ref{hatRdef})
for the Kerr-Schild metrics defined
by (\ref{1rot1}) and (\ref{1rot2}) can be given as follows.  With
$x^a=(t,r,\theta,\phi)$ and $x^i$ being the coordinates of the
$(2n-3)$-sphere, the non-vanishing components of
$\wR^{\mu\nu}_{\rho\sigma}$ involve either four, two (one up, one down)
or zero $(2n-3)$-sphere
indices.  Writing $c\equiv \cos\theta$ and $s\equiv \sin\theta$, we find
\be
\wR^{ij}_{kl} = \fft{\lambda a^2 c^2}{r^2} \delta_{kl}^{ij}\,,\qquad
\wR^{ai}_{bj} = T^a_b\, \delta^i_j\,,\label{fres}
\ee
where
\bea
&&
T^{t}_{t}=\fft{2\lambda a^2 s^2\,\Delta_\theta^2}{\Xi_a
   (g^2 r^2 + 1)\rho^2}\,,\qquad
T^{\phi}_{t}= \fft{2\lambda a^3 s^2\,\Delta_\theta^2}{\Xi_a
(r^2 + a^2)\rho^2}\,,\qquad
T^{\theta}_{t}=\fft{2\lambda a^2 c s \Delta_\theta^2}{\Xi_a r \rho^2}\,,\nn\\
&&
T^{r}_{t} = \fft{2\lambda a^2 s^2 \Delta_\theta^2}{\Xi_a \rho^2}\,,\qquad
T^{t}_{\phi}=-\fft{2\lambda a^3 s^4\,
    \Delta_\theta}{\Xi_a(g^2r^2+1) \rho^2}\,,\qquad
T^{\phi}_{\phi} = -\fft{2\lambda a^4 s^4\,
\Delta_\theta}{\Xi_a(r^2+a^2) \rho^2}\,,\nn\\
&&
T^{\theta}_{\phi}= -\fft{2\lambda a^3 c s^3\theta\,
\Delta_{\theta}}{\Xi_a r \rho^2}\,,\qquad
T^{r}_{\phi}= -\fft{2\lambda a^3 s^4\Delta_{\theta}}{\Xi_a\rho^2}\,,\qquad
T^{t}_{\theta} = - \fft{2\lambda a^2 c s}{r(g^2 r^2 + 1)}\,,\nn\\
&&
T^{\phi}_{\theta} = - \fft{2\lambda a^3 c s}{r(r^2 + a^2)}\,,\qquad
T^{r}_{\theta} = -\fft{2\lambda a^2 c s}{r}\,,\qquad
T^{t}_{r} = - \fft{2\lambda a^2 s^2\Delta_{\theta}}{(g^2r^2 +1)^2
       (r^2 + a^2)}\,,\nn\\
&&
T^{\phi}_{r} = - \fft{2\lambda a^3 s^2\Delta_{\theta}}{(g^2r^2 +1)
   (r^2 + a^2)^2}\,,\qquad
T^{\theta}_{r} = -\fft{2\lambda a^2 c s\Delta_{\theta}}{r(g^2r^2+1)
                     (r^2+a^2)}\,,\nn\\
&&
T^{r}_{r} = - \fft{2\lambda a^2 s^2\Delta_{\theta}}{(g^2r^2 + 1)
       (r^2 + a^2)}\,,\qquad
T^\theta_\theta=0\,,\label{Tab}
\eea
and the components $\wR^{ab}_{cd}$ are given by
\bea
&&\wR^{t\,\phi}_{t\,\phi}=\fft{2\lambda a^2 c^2[-(r^2 + a^2)(\Xi_a+2a^2g^2 s^2) + a^4 g^2 s^4]}{(g^2r^2 + 1)(r^2 + a^2) \rho^2}\,,\quad
\wR^{t\,\phi}_{t\,\theta} = -\fft{2\lambda a^3 g^2 r cs}{(g^2r^2+1)(r^2+a^2)}\,,\nn\\
&&
\wR^{t\,\phi}_{t\,r} = \fft{2\lambda a^5 g^2 c^2 s^2}{(g^2r^2+1)
   (r^2+a^2)^2}\,,\qquad
\wR^{t\,\phi}_{\phi\,\theta} = \fft{2\lambda a^2 r c s}{(g^2r^2 + 1)(r^2 + a^2)}\,,\nn\\
&&
\wR^{t\,\phi}_{\phi\,r} =-\fft{2\lambda a^2 c^2
  \Delta_{\theta}}{(g^2r^2+1)^2(r^2 + a^2)}\,,\qquad
\wR^{t\,\phi}_{\theta\,r} = \fft{2\lambda a^3 r c s \Xi_a}{(g^2 r^2 + 1)^2
(r^2 + a^2)^2}\,,\nn\\
&&
\wR^{t\,\theta}_{t,\phi} = -\fft{2\lambda a^3 g^2 r c s^3
\Delta_{\theta}}{\Xi_a (g^2 r^2+1)\rho^2}\,,\qquad
\wR^{t\,\theta}_{t\,r} = -\fft{r s \Delta_{\theta}}{\Xi_a c \rho^2}
  \wR^{t\,\phi}_{t\,\phi}\,,\qquad
\wR^{t\,r}_{t\,\phi}=\fft{2\lambda a^5 g^2 c^2 s^4}{\Xi_a\rho^2}
\,,\nn\\
&&
\wR^{t\,\theta}_{t\,\theta} = \fft{2\lambda a^2}{\Xi_a (g^2 r^2 + 1)
   \rho^4}\Big[r^2 + r^2 (2a^2 g^4 r^2 - 5 a^2 g^2 -2) c^2 \nn\\
&&\qquad\qquad\qquad\qquad\qquad- a^2(2g^4 r^4 - 8 a^2 g^4 r^2 - 7 g^2 r^2 +3a^2 g^2 +1) c^4\nn\\
&&\qquad\qquad\qquad\qquad\qquad  - a^4 g^2( 9g^2 r^2 - 5a^2 g^2 -5) c^6 - 6a^6 g^4 c^8\Big]\,,\nn\\
&&
\wR^{t\,\theta}_{\phi\,\theta}=-\fft{2\lambda a^3 s^2\Big[
r^2+ r^2(2g^2 r^2 - 3 a^2 g^2 -3) c^2 + a^2 (9 g^2 r^2 -2 a^2 g^2 -2) c^4 + 6a^4 g^2 c^6\Big]}{\Xi_a (g^2 r^2 + 1)\rho^4},\nn\\
&&
\wR^{t\,\theta}_{\phi\,r} = -\fft{2\lambda a^3 r c s^3
\Delta_\theta(2 g^2 r^2 + a^2 g^2 c^2 + 1)}{\Xi_a
      (g^2 r^2 + 1)^2 \rho^4}\,,\qquad
\wR^{t\,r}_{t\,\theta}=\fft{(g^2 r^2 + 1)(r^2 + a^2)}{\Delta_\theta}
\wR^{t\,\theta}_{t\,r}\,,\nn\\
&&
\wR^{t\,\theta}_{\theta\,r}=\fft{2\lambda a^2(1 - (5a^2 g^2 +2) c^2 +
  6 a^2 g^2 c^4)}{
(g^2 r^2 + 1)^2 (r^2 + a^2)},\qquad \wR^{t\,r}_{\phi\,r} = -\fft{2\lambda a^5 c^2 s^4 \Delta_\theta}{
\Xi_a (g^2 r^2 + 1) \rho^4}\,,\nn\\
&&
\wR^{t\,r}_{\phi\,\theta} = \fft{2\lambda a^3 cs^3
   r[a^2 s^2 - 2(r^2 + a^2)]}{
\Xi_a\rho^4}\,,\qquad
\wR^{t\,r}_{t\,r} = \fft{2\lambda a^4 c^2 s^2
[(\Xi_a (r^2 + a^2) + a^4 g^2 s^4]}{\Xi_a(r^2 + a^2) \rho^2}\,,\nn\\
&&
\wR^{t\,r}_{\theta\,r} = \fft{2\lambda
                      a^2 c s r}{(g^2 r^2 +1)(r^2 + a^2)}\,,\quad
\wR^{\phi\,\theta}_{t\,\phi} = -\fft{2\lambda a^2 c s
             \Delta_{\theta}^2 r}{\Xi_a (r^2 + a^2) \rho^2}\,,\quad
\wR^{\phi\,\theta}_{t\,\theta} = -\fft{(g^2r^2 + 1)
    \Delta_{\theta}}{(r^2 +a^2)s^2}
\wR^{t\,\theta}_{\phi\,\theta}\,,\nn\\
&&
\wR^{\phi\,\theta}_{t\,r}=-\fft{\Delta_\theta^2}{(r^2 + a^2)^2 s^2}
     \wR_{\phi\,\theta}^{t\,r}\,,\qquad
\wR^{\theta\,r}_{t\,\phi} = -\fft{2\lambda a^3 c s^3
  \Delta_\theta^2 r}{\Xi_a\rho^4}\,,\nn\\
&&
\wR^{\phi\,\theta}_{\phi\,\theta}=\fft{2\lambda a^2}{\Xi_a(r^2 + a^2)
   \rho^4}\Big[
6a^6 g^2 c^8 -a^4 (5 + 5 a^2 g^2 - 9 g^2 r^2) c^6 - a^2 r^2\nn\\
&&\qquad\qquad
+a^2 (3 a^2 + a^4 g^2 - 8 r^2 - 7 a^2 g^2 r^2 + 2 g^2 r^4) c^4 +
r^2 (5 a^2 + 2 a^4 g^2 - 2 r^2) c^2\Big]\,,\nn\\
&&
\wR^{\phi\,\theta}_{\phi\,r}=\fft{s\Delta_\theta r}{\Xi_a \rho^2 c}
   \wR^{t\,\phi}_{t\,\phi}\,,\qquad
\wR_{\phi\,\theta}^{\phi\,r}=\fft{(g^2 r^2 + 1)(r^2 + a^2)}{\Delta_\theta}
   \wR^{\phi\,\theta}_{\phi\,r}\,,\nn\\
&&
\wR^{\phi\,\theta}_{\theta\,r}=\fft{2\lambda a^3(1  -
   c^2 (5 + 2 a^2 g^2) +6 a^2 c^4 g^2)}{
(g^2r^2 + 1)(r^2 + a^2)^2}\,,\qquad
\wR_{\phi\,\theta}^{\theta\,r}=\fft{(g^2r^2+1)(r^2 + a^2)^2s^2}{\Xi_a
\rho^2} \wR^{\phi\,\theta}_{\theta\,r}\,,\nn\\
&&
\wR^{\phi\,r}_{t\,\phi}=-\fft{(g^2r^2+1)^2 (r^2 + a^2)\Delta_\theta}{\Xi_a
\rho^2} \wR_{\phi\,r}^{t\,\phi}\,,\qquad
\wR^{\phi\,r}_{t\,\theta}=-\fft{(g^2r^2+1)^2}{s^2} \wR_{\phi\,r}^{t\,\theta}\,,
\label{Rabcd}\\
&&
\wR^{\phi\,r}_{t\,r}=-\fft{(g^2r^2+1) \Delta_\theta}{(r^2+a^2) s^2}
   \wR_{\phi\,r}^{t\,r}\,,\qquad
\wR^{\phi\,r}_{\phi\,r} = \fft{2\lambda a^2 c^2 \Delta_\theta[\Xi_a
 (r^2 + a^2-2a^2s^2) - a^4 g^2 s^4]}{\Xi_a (g^2r^2 + 1)\rho^4}\,,\nn\\
&&
\wR^{\phi\,r}_{\theta\,r}=\fft{2\lambda a^3g^2 c s r}{(g^2r^2 + 1)
     (r^2 + a^2)}\,,\qquad
\wR_{\phi\,r}^{\theta\,r}=\fft{g^2a(r^2 + a^2) s^2 \Delta_\theta}{\Xi_a \rho^2}
  \wR^{\phi\,r}_{\theta\,r}\,,\nn\\
&&
\wR^{\theta\,r}_{t\,\theta}=-\fft{(g^2r^2+1)^2 (r^2 + a^2)
\Delta_\theta}{\Xi_a\rho} \wR_{\theta\,r}^{t\,\theta}\,,\qquad
\wR^{\theta\,r}_{t\,r}=-\fft{(g^2r^2+1)\Delta_\theta^2}{\Xi_a\rho^2}
\wR_{\theta\,r}^{t\,r}\,,\nn\\
&&
\wR^{\theta\,r}_{\theta\,r}=-\fft{2\lambda a^2[r^2-2 (-2 a^2 + r^2 +
   a^2 g^2 r^2)c^2
 - a^2 (5 + 5 a^2 g^2 - 3 g^2 r^2)c^4 +6 a^4  g^2 c^6]}{(g^2r^2 + 1)
 (r^2 + a^2) \rho^2}\,.\nn
\eea

\end{document}